\documentclass[12pt]{article}
 \newcommand{\be}{\begin{equation}}
 \newcommand{\ee}{\end{equation}}
 \newcommand{\bea}{\begin{eqnarray}}
 \newcommand{\eea}{\end{eqnarray}}
 \newcommand{\nn}{\nonumber}

 \date{October 29, 2003}
\title{Complete description of polarization effects in the nonlinear Compton
scattering \\
 {\it I. Circularly polarized laser photons}}

\author{D.Yu.~Ivanov$^{1)}$, G.L.~Kotkin$^{2)}$,
V.G.~Serbo$^{2)}$\\ {\it $^{1)}$Sobolev Institute of Mathematics,
Novosibirsk, 6300090, Russia} \\ {\it $^{2)}$Novosibisk State
University, Novosibirsk, 6300090, Russia } }

\begin{document}
 \maketitle

 \begin{abstract}

 We consider emission of a photon by an electron in the field of a
 strong laser wave. Polarization effects in this process are
 important for a number of physical problems. We discuss
 a probability of this process
 for circularly polarized laser photons and for arbitrary
 polarization of all other particles. We
 obtain the complete set of functions which describe such a
 probability in a compact covariant form. Besides, we discuss an
 application of the obtained formulas to the problem of $e\to
 \gamma$ conversion at $\gamma \gamma$ and $\gamma e$
 colliders.

 \end{abstract}

\section{Introduction}

The Compton scattering
 \be
  e(p) +\gamma (k) \to e(p') +\gamma (k')
  \label{1}
 \ee
is one of the first processes calculated in quantum
electrodynamics at the end of 20's. The analysis of its
polarization effects is now included in text-books (see, for
example, \cite{BLP} \S 87). Nevertheless, the complete description
of cross sections with the polarization of both initial and final
particles has been considered in detail only recently (see
\cite{Grozin,GKPS83,KPS98} and literature therein). One of the
interesting application of this process is related to collisions
of ultra-relativistic electrons with the beam of polarized laser
photons. In this case the Compton effect is a basic process at
obtaining high-energy photons for contemporary experiments in
nuclear physics (photonuclear reactions with polarized photons)
and for future $\gamma \gamma $ and $\gamma e$ colliders
\cite{GKST83}. The importance of the particle polarization is
clearly seen from the fact that the number of photons with maximum
energy is nearly doubled when helicities of the initial electron
and photon are opposite.

With the growth of the laser field intensity, an electron collides
(with an essential probability) with $n$ laser photons
simultaneously,
 \be
   e(q) +n\,\gamma_L (k) \to e(q') +\gamma (k')\,,
 \label{2}
 \ee
thus the Compton scattering becomes nonlinear. Such a process with
absorbtion of $n=1,\, 2,\, 3,\, 4$ laser photons was observed in
recent experiment at SLAC~\cite{SLAC}. The polarization properties
of process (\ref{2}) are important for a number of problems, for
example, for obtaining highly polarized electrons and positrons
beams~\cite{Tsai}, for a laser beam cooling~\cite{Telnov} and,
especially, for future $\gamma \gamma $ and $\gamma e$ colliders.
In the latter case the nonlinear Compton scattering must be taken
into account at simulation of the processes in the conversion
region. For comprehensive simulation, including processes of
multiple electron scattering, one has to know not only the
differential cross section of the nonlinear Compton scattering,
but energy, angles and polarization of final photons and electrons
as well. The method for calculation of such cross sections was
developed by Nikishov and Ritus~\cite{NR-review}. It is based on
the exact solution of the Dirac equation in the external
electromagnetic  plane wave. Some particular polarization
properties of this process for the circularly polarized laser
photons were considered in~\cite{GS,Tsai,Yokoya,BPM} and have
already been included in the existing simulation
codes~\cite{Yokoya,TCode}. In the present paper we give the
complete description of the nonlinear Compton scattering for the
case of circularly polarized laser photons and arbitrary
polarization of all other particles. The case of the linearly
polarized laser photons will be considered in the next paper. We
follow the method of Nikishov and Ritus in the form presented
in~\cite{BLP} \S 101.

In the next section we describe the kinematics. The cross section
in the invariant form, including polarization of all particles, is
obtained in Sect. 3. In Sect. 4 we considered in detail the
polarization of final particles in the frame of reference relevant
for $\gamma \gamma$ and $\gamma e$ colliders. In Sect. 5 we
summarized  our results and compare them with those known in the
literature. In Appendix we give a comparison of the obtained cross
section in the limit of weak laser field with the cross section
for the linear Compton scattering.

\section{Kinematics}

\subsection{Parameter of nonlinearity}

Let us consider the interaction of an electron with a
monochromatic plane wave described  by 4-potential $A_\mu$ (the
corresponding electric and magnetic fields are ${\bf E}$ and ${\bf
B}$, a frequency is $\omega$) and let $F$ be the root-mean-squared
field strength,
 $$
 F^2 =\langle {\bf B}^2 \rangle = \langle {\bf E}^2 \rangle\,.
  $$
The invariant parameter describing the intencity of the laser
field (the parameter of nonlinearity) is defined via the mean
value of squared 4-potential:
 \be
\xi={e\over mc^2} \sqrt{-\langle A_\mu A^\mu \rangle}\,,
 \label{3}
 \ee
where $e$ and $m$ is the electron charge and mass, $c$ is the
velocity of light. The origin of this parameter can be explained
as follows. The electron oscillates in the transverse direction
under influence of the force $\sim eF$ and for the time $\sim 1/
\omega$ acquires the transverse momentum $\sim p_\perp= eF /
\omega$, thus for the longitudinal motion its effective mass
becomes $m_*= \sqrt{m^2+(p_\perp/c)^2}$. The ratio of the momentum
$p_\perp$ to $mc$ is the natural dimensionless parameter
 \be
 \xi = {eF\over m\,c\,\omega}\,.
 \label{4}
  \ee
This parameter can be also expressed via the density $n_L$ of
photons in the laser wave:
 \be
\xi^2 = \left({e F \over m \omega c}\right)^2={4\pi \alpha
\hbar\over m^2c\,\omega}\, n_L\,,
 \label{5}
  \ee
where $\hbar$ is the Plank constant.

>From the classical point of view, the oscillated electron emits
harmonics with frequencies $n\,\omega$, where $n=1,\;2,\;\dots$.
Their intensities at small $\xi$ is proportional to $({\bf E}^2)^n
\propto \xi^{2n}$, the polarization properties of these harmonics
depends on the polarizations of the laser wave and the initial
electron\footnote{The polarization of the first harmonic is
related to the tensor of the second rank $\langle
E_iE^*_j\rangle$, in this case one needs only three Stokes
parameters. The polarization properties of higher harmonics are
connected with the tensors of higher rank $\langle E_iE_j \dots
E^*_k\rangle$, in this case one needs more parameters for their
description. It is one of the reason why the nonlinear Compton
effect had been considered only for 100\% polarized laser beam
mainly for circular or linear polarization.}. From the quantum
point of view, this radiation can be described as the nonlinear
Compton scattering with absorbtion of $n$ laser photons. When
describing such a scattering, we have to take into account that in
a laser wave 4-momenta $p$ and $p'$ of the free initial and final
electrons are replaced by the 4-quasi-momenta $q$ and $q'$
(similar to the description of the particle motion in a periodic
potential field in non-relativistic quantum mechanics),
 \bea
q&=&p+\xi^2{ m^2c^2\over 2pk}\,k\,,\;\; q'=p'+\xi^2 {m^2c^2\over
2p'k}\,k\,,
 \label{6}\\
 q^2&=&(q')^2= (1+\xi^2 )\,m^2c^2\equiv
m_*^2c^2\,.
 \nn
 \eea
In particular, the energy of the free incident electron $E$ is
replaced by the quasi-energy
 \be
c q_0=E+\xi^2 {m^2c^2\over 2pk}\,\hbar\omega\,.
 \label{7}
 \ee

\subsection{Invariant variables}

As a result, we deal with the reaction (\ref{2}) for which the
conservation law reads
 \be
 q+n\,k=q'+k'\,.
 \label{8}
 \ee
From this it follows that all the kinematic relations, which occur
for the linear Compton scattering, will apply to the process
considered here if the electron momenta $p$ and $p'$ are replaced
by the quasi-momenta $q$ and $q'$ and the incident photon momentum
$k$ by the 4-vector $n k$. Since $qk=pk$, we can use the same
invariant variables  as for the linear Compton scattering (compare
\cite{GKPS83}):
 \be
x={2p k\over m^2c^2}\,, \;\;\; y\; = {k k'\over p k}\,.
 \label{9}
 \ee
Moreover, many kinematic relations can be obtained from the
previous ones by the replacement: $\omega \to n\, \omega$, $m\to
m_*$. In particular, we introduce convenient combinations
analogous to those used in the linear Compton scattering
\begin{equation}
s_n=2\sqrt{r_n(1-r_n)},\;\; c_n= 1-2r_n\,,
  \label{10}
\end{equation}
where
 \be
 r_n={y\,(1+\xi^2)\over (1-y)\,nx}\,.
 \label{11}
 \ee
It is useful to note that these invariants have a simple notion,
namely
 \be
s_n=\sin{\tilde\theta},\;\; c_n=\cos{\tilde\theta}\,,\;\;
r_n=\sin^2(\tilde \theta/2)\,,
 \label{12}
 \ee
where $\tilde\theta$ is the photon scattering angle in the frame
of reference where the initial electron is at rest on average
(${\bf q}=0$). Therefore,
 \be
0 \leq s_n,\; r_n \leq 1\,,\;\;-1 \leq c_n \leq 1\,.
 \label{13}
 \ee

The maximum value of the variable $y$ for the reaction (\ref{2})
is
 \begin{equation}
 y\leq \;y_n\;= {nx\over nx+1+\xi^2}\,.
\label{14}
\end{equation}
The value of $y_n$ is close to $1$ for large $n$, but for a given
$n$ it decreases with the growth of the parameter $\xi$. With this
notation one can rewrite $s_n$ in another form
 \be
  s_n= {2\over y_n(1-y)}\, \sqrt{y(y_n-y)(1-y_n)}\,,
 \label{15}
 \ee
from which it follows that
 \be
 s_n \to 0 \;\;\mbox{ at } \;\;y\to y_n \;\; \mbox{or at} \;\;
 y\to0\,.
 \label{16}
 \ee

The usual notion of the cross section is not applicable for
reaction (\ref{2}) and the description of this reaction is usually
given in terms of the probability of the process per second
$\dot{W}$. However, for the procedure of simulation in the
conversion region as well as for the simple comparison with the
linear case, it is useful to introduce the ``effective cross
section'' given by the definition
 \be
d\sigma= {d\dot{W}\over j}\,,
 \label{17}
 \ee
where
 $$
j= {(q\,nk) c^2\over q_0\, n \hbar\omega}\,n_L ={m^2c^4 \,x\over
2q_0 \hbar\omega} \, n_L
 $$
is the flux density of colliding particles. Contrary to the usual
cross section, this effective cross section does depend on the
laser beam intensity, i.e. on the parameter  $\xi$. This cross
section is the sum over harmonics, corresponding to the reaction
(\ref{2}) with a given number $n$ of the absorbed laser
photons\footnote{In this formula and below the sum is over those
$n$ which satisfy the condition $y< y_n$, i.e. this sum runs from
some minimal value $n_{\min}$ up to $n=\infty$, where $n_{\min}$
is determined by the equation $y_{n_{\min}-1} < y <
y_{n_{\min}}$.}:
  \be
d\sigma=\sum\limits_{n}d\sigma_n\,.
 \label{18}
 \ee

\subsection{Invariant polarization parameters}

The invariant description of the polarization properties of both
the initial and the final photons can be performed in the standard
way (see \cite{BLP} \S 87). We define a pair of unit 4-vectors
\begin{equation}
e^{(1)}={N\over\sqrt{-N^2}}\,,\;\;\;
e^{(2)}={P\over\sqrt{-P^2}}\,,
 \label{19}
\end{equation}
where\footnote{Below we use the system of units in which $c=1$,
$\hbar=1$.}
 \bea
N^\mu&=&\varepsilon^{\mu\alpha \beta \gamma} P_\alpha
(k'-n\,k)_\beta K_\gamma \, ,\;\;\;
P_\alpha=(q+q')_\alpha-{(q+q')K\over K^2}\,K_\alpha \,,\;\;\;
K_\alpha=n\,k_\alpha+k'_\alpha\,,
 \nn\\
\sqrt{-N^2}&=&m^3xy\, {s_n\over r_n} \sqrt{1+\xi^2}\,,\;\;\;
\sqrt{-P^2}=m \, {s_n\over r_n} \sqrt{1+\xi^2}\,.
 \nn
 \eea
The 4-vectors $e^{(1)}$ and $e^{(2)}$ are orthogonal to each other
and to 4-vectors $k$ and $k'$,
 $$
e^{(i)} e^{(j)} = - \delta_{ij}, \;\; e^{(i)} k = e^{(i)} k' =0;
\;\; i,\, j = 1,\,2 \,.
 $$
Therefore, they are fixed with respect to the scattering plane of
the process.

Let $\xi_j$ and $\xi '_j$ be the Stokes parameters for the initial
and final photon which are defined with respect to 4-vectors
$e^{(1)}$ and $e^{(2)}$. In the considered case of 100\%
circularly polarized laser beam
 \be
\xi_1=\xi_3=0,\;\; \xi_2=P_c= \pm 1\,,
 \label{20}
 \ee
where $P_c$ is the degree of the circular polarization of the
laser wave or the initial photon helicity. The parameters $\xi
'_j$ enter the Compton cross section, they describe {\it the
detector polarization} which essentially represents the properties
of the detector as selecting one or other polarization of the
final photon.

Let {\boldmath $\zeta$} and {\boldmath $\zeta $}$^{\prime}$ be the
polarization vectors of the initial and final electrons. They
determine the electron-spin 4-vectors
\begin{equation}
a=\left({\mbox {\boldmath $\zeta$}{\bf p}\over m}\,,\,
\mbox{\boldmath $\zeta$}+ {\bf p}{\mbox{\boldmath $\zeta$} {\bf p}
\over m(E+m)}\right) \,,\;\;\; a^{\prime}=\left({\mbox
{\boldmath$\zeta$}^{\prime} {\bf p}^{\prime}\over m}\,,\, \mbox
{\boldmath$\zeta$}^{\prime} +{\bf p}^{\prime}{\mbox
{\boldmath$\zeta$}^{\prime} {\bf p}^{\prime}\over
m(E^{\prime}+m)}\right)
 \label{21}
\end{equation}
and the mean helicity of the initial and final electrons
\begin{equation}
\lambda_e={\mbox{\boldmath $\zeta$}{\bf p}\over 2|{\bf
p}|}\,,\;\;\;\; \lambda^{\prime}_e={\mbox {\boldmath$
\zeta$}^{\prime} {\bf p}^{\prime}\over 2|{\bf p}^{\prime}|}\,.
 \label{22}
\end{equation}
Now we have to define invariants, which describe the polarization
properties of the initial and the final electrons. For the
electrons it is a more complicated task than it is for the
photons. For the linear Compton scattering, the relatively simple
description was obtained in ~\cite{Grozin} using invariants which
have a simple meaning in the centre-of-mass system. However, this
frame of reference is not convenient for the description of the
nonlinear Compton scattering since it has actually to vary with
the change of the number of laser photons $n$.

Our choice is based on the experience obtained in~\cite{KPS98} and
\cite{BPM}. We exploit two ideas. First of all, the 4-vector
$e^{(1)}$ is orthogonal to 4-vectors $k,\;k',\; p$ and $p'$,
therefore, the invariants $\zeta_1 =-ae^{(1)}$ and $\zeta'_1
=-a'e^{(1)}$ are the transverse polarizations of the initial and
the final electrons perpendicular to the scattering plane.
Further, it is not difficult to check that invariant $ak/(2mx)$ is
the helicity of the initial electron in the frame of reference, in
which the electron momentum ${\bf p}$ is anti-parallel to the
initial photon momentum ${\bf k}$. Analogously, the invariant
$a'k/[(2mx(1-y)]$ is the helicity of the final electron in the
frame of reference, in which momentum of the final electron ${\bf
p}'$ is anti-parallel to the initial photon momentum ${\bf k}$. It
is important that this interpretation is valid for any number of
the absorbed laser photons. Furthermore, we will show that for
practically important case the latter frame of reference is almost
coincides with the former frame of reference.

Taken into account that $ap=a'p'=0$, we find two sets of units
4-vectors:
 \bea
e_1&=&e^{(1)},\;\;\; e_2=-e^{(2)}-{\sqrt{-P^2}\over m^2 x}k,\;\;\;
e_3={1\over m}\left(p-{2\over x}k\right)\,;
 \label{23}\\
e^{\prime}_1&=&e^{(1)},\;\;\;e^{\prime}_2=-e^{(2)}-{\sqrt{-P^2}
\over m^2 x(1-y)}k,\;\;\; e^{\prime}_3={1\over
m}\left(p^{\prime}-{2\over x(1-y)}k\right)\,.
 \nn
 \eea
These vectors satisfy the conditions:
 \begin{equation}
e_i e_j=-\delta_{ij}\;,\;\;\;e_j p=0\; ; \;\;\;e^{\prime}_i
e^{\prime}_j= -\delta_{ij}\; , \;\;\; e^{\prime}_j p^{\prime}=0\,.
 \label{24}
\end{equation}

It allows us to represent the 4-vectors $a$ and $a^{\prime}$ in
the following covariant form
\begin{equation}
a=\sum_{j=1}^3 \zeta_j e_j\,,\;\;\; a^{\prime}=\sum_{j=1}^3
\zeta^{\prime}_j e^{\prime}_j \,,
 \label{25}
\end{equation}
where
 \be
 \zeta_j=-ae_j\,,\;\;\;
\zeta^{\prime}_j=-a^{\prime}e^{\prime}_j\,.
 \label{26}
\end{equation}
The invariants $\zeta_j$ and $\zeta^{\prime}_j$ describe
completely the polarization properties of the initial electron and
the detector polarization properties of the final electron,
respectively. To clarify the meaning of invariants $\zeta_j$, it
is useful to note that
 \begin{equation}
\zeta_j ={\mbox{\boldmath $\zeta$}}{\bf n}_j\,,
 \label{27}
\end{equation}
where the corresponding 3-vectors are
\begin{equation}
{\bf n}_j ={\bf e}_j -{{\bf p}\over E+m}\;e_{j0}
 \label{28}
\end{equation}
with $e_{j0}$ being a time component of 4-vector $e_{j}$ defined
in (\ref{23}).  Using properties (\ref{24}) for 4-vectors $e_j$,
one can check that
\begin{equation}
{\bf n}_i \; {\bf n}_j =\delta_{ij}\,.
 \label{29}
\end{equation}
As a result, the polarization vector  $\mbox{\boldmath$\zeta$}$
has the form
\begin{equation}
\mbox{\boldmath$\zeta$}=\sum^3_{j=1} \zeta_j\;{\bf n}_j\, .
 \label{30}
\end{equation}
Analogously, for the final electron the invariants
$\zeta^{\prime}_j$ can be presented as
\begin{equation}
\zeta'_j ={\mbox{\boldmath $\zeta$}}'{\bf n}'_j, \;\;\; {\bf n}'_j
={\bf e}'_j -{{\bf p}'\over E'+m}\;e'_{j0}, \;\;\; {\bf n}'_i \;
{\bf n}'_j =\delta_{ij}\,,
 \label{31}
\end{equation}
such that
\begin{equation}
\mbox{\boldmath$\zeta$}'=\sum^3_{j=1} \zeta'_j\;{\bf n}'_j\, .
 \label{32}
\end{equation}

\section{Cross section in the invariant form}

\subsection{General relations}

The effective differential cross section can be presented in the
following invariant form
\begin{equation}
d\sigma (\xi^{\prime}_i,\zeta^{\prime}_j) = {r_e^2\over
4x}\;\sum_n F^{(n)}\;d\Gamma_n\,,\;\;\;d\Gamma_n = \delta
(q+n\,k-q-k')\;{d^3k'\over \omega'}{d^3q'\over q_0'}\,,
 \label{33}
 \end{equation}
where $r_e=\alpha/m$ is the classical electron radius, and
\begin{equation}
F^{(n)}=F_0^{(n)}+\sum ^3_{j=1}\left( F_j^{(n)}\xi '_j\; +
\;G_j^{(n)} \zeta^{\prime}_j\right) + \sum
^3_{i,j=1}H_{ij}^{(n)}\,\zeta^{\prime}_i\,\xi^{\prime}_j \,.
 \label{34}
\end{equation}
Here function $F_0^{(n)}$ describes the total cross section for a
given harmonic $n$, summed over spin states of the final
particles:
 \be
\sigma_n = {r_e^2\over x}\; \int F_0^{(n)}\,d\Gamma_n \,.
 \label{35}
 \ee
Items $F_j^{(n)}\xi '_j$  and $G_j^{(n)} \zeta^{\prime}_j$ in Eq.
(\ref{34}) describe the polarization of the final photons and the
final electrons, respectively. The last items $H_{ij}^{(n)}
\zeta^{\prime}_i\,\xi^{\prime}_j$ stand for the correlation of the
final particles' polarizations.

>From Eqs. (\ref{33}), (\ref{34}) it is possible to deduce {\it the
polarization of the final photon and electron resulting from the
scattering process itself}. We denote the Stokes parameters
describing this polarization by $\xi_j^{(f)}$ to distinguish them
from the detected polarization $\xi'_j$. Analogously, we denote by
$\zeta^{(f)}_j$ the invariant parameters of this polarization for
the final electron to distinguish them from the detector
polarization parameters $\zeta'_j$. According to the usual rules
(see~\cite{BLP} \S 65), we obtain the following expression for the
Stokes parameters of the final photon:
\begin{equation}
\xi_j^{(f)}= {F_j\over F_0}\,,\;\; F_0=\sum_n
F_0^{(n)}\,,\;\;F_j=\sum_n F_j^{(n)}\,;\;\; j= 1,\,2,\,3\,.
 \label{36}
\end{equation}
The polarization of the final electron is given by invariants
 \be
\zeta_j^{(f)}= {G_j\over F_0}\,,\;\;G_j=\sum_n G_j^{(n)}\,,
 \label{37}
 \ee
therefore, its polarization vector is
 \be
\mbox{ \boldmath $\zeta$}^{(f)}= \sum_{j=1}^3 \, {G_j\over F_0}\,
{\bf n}_j^{\prime}\,.
 \label{38}
 \ee
In the similar way, the polarization properties for a given
harmonic $n$ are described by:
\begin{equation}
\xi_{j\,(n)}^{(f)}= {F_j^{(n)}\over F_0^{(n)}}\,,\;\;
\zeta_{j\,(n)}^{(f)}= {G_j^{(n)}\over F_0^{(n)}}\,.
 \label{39}
\end{equation}

\subsection{The results}

We have calculated coefficients $F_j^{(n)},\, G_j^{(n)}$ and
$H_{ij}^{(n)}$ using the standard technic presented in~\cite{BLP}
\S 101. All the necessary traces have been calculated using the
package MATE\-MA\-TIKA. In the considered case of 100 \%
circularly polarized ($P_c=\pm 1$) laser beam, almost all
dependence on the parameter $\xi$ accumulates in three functions:
  \bea
f_n& \equiv&
f_n(z_n)=J_{n-1}^{2}(z_n)+J_{n+1}^{2}(z_n)-2J_{n}^{2}(z_n)\,,
 \nn\\
 g_n& \equiv& g_n(z_n)=\frac{4 n^2 J_{n}^2(z_n)}{z_n^2}\,,
 \label{40}\\
h_n&\equiv& h_n(z_n)=J_{n-1}^{2}(z_n)-J_{n+1}^{2}(z_n)\,,
 \nn
 \eea
where $J_n(z)$ is the Bessel function. Functions (\ref{40}) depend
on $x$, $y$ and $\xi$ via the single argument
 \be
z_n= {\xi\over \sqrt{1+\xi^2}}\; n\,s_n\,.
 \label{41}
 \ee
For small value of this argument one has
\be
f_n=g_n=h_n=\frac{(z_n/2)^{2(n-1)}}{[(n-1)!]^2} \;\; \mbox{ at}
\;\; z_n \to 0\,,
 \label{42}
 \ee
in particular,
 \be
 f_1=g_1=h_1=1
\;\; \mbox{ at} \;\; z_1= 0\,.
 \label{43}
 \ee
It is useful to note that this argument is small for small $\xi$,
as well as for small or for large values of $y$:
 \be
 z_n \to 0 \;\; \mbox{either at} \;\; \xi\to 0\,, \;
  \mbox{ or at }\;\; y\to y_n\,, \;\;\mbox{ or at }\;\; y\to 0\,.
  \label{44}
 \ee

The results of our calculations are the following. The item
$F_0^{(n)}$, related to the total cross section (\ref{35}), reads
 \bea
F_0^{(n)}&=&\left({1\over 1-y}+1-y\right)\,f_n- {s_n^2\over
1+\xi^2}\, g_n-
 \nn
 \\
&& \left[{y s_n \over \sqrt{1+\xi^2}}\, \zeta_2 -{y(2-y)\over
1-y}\, c_n \,\zeta_3\right] \,h_n\,P_c\,.
 \label{45}
 \eea

The polarization of the final photons $\xi_j^{(f)}$ is given by
Eq. (\ref{36}) where
  \bea
F_1^{(n)}&=&{y\over 1-y} {s_n \over \sqrt{1+\xi^2}}\, h_n P_c\,
\zeta_1\,,
 \label{46}
 \\
F_2^{(n)}&=&\left({1\over 1-y} +1-y\right)\, c_n h_n P_c- {ys_n
c_n \over \sqrt{1+\xi^2}}\, g_n \, \zeta_2+
 \nn
 \\
&&y\left( {2-y\over 1-y}\,f_n- {s^2_n\over 1+\xi^2}\,g_n\right)
\zeta_3\,,
 \nn
 \\
F_3^{(n)}&=&2(f_n-g_n)+ s^2_n (1+\Delta)\,g_n -{y\over 1-y} {s_n
\over \sqrt{1+\xi^2}}\, h_n P_c\, \zeta_2 \,,
 \nn
 \eea
here we use the notation
 $$
 \Delta= {\xi^2 \over 1+\xi^2}\,.
 $$

The polarization of the final electrons $\zeta_j^{(f)}$ is given
by Eqs. (\ref{37}), (\ref{38}) with
 \bea
G_1^{(n)}&=&\, \langle G_1^{(n)} \rangle\, \zeta_1,
 \label{47}\\
G_2^{(n)}&=&\, \langle G_1^{(n)} \rangle\, \,\zeta_2 -
\frac{y{s}_n}{(1-y)\sqrt{1+\xi^2}}\,\left(c_n\,g_n\, \zeta_3+
h_n\, P_c\right)\,,
 \nn\\
G_3^{(n)}&=&\, \langle G_3^{(n)} \rangle\,+\, \frac{y {s}_n
{c}_n}{\sqrt{1+\xi^2}}\,g_n \,\zeta_2,
 \nn
  \eea
where we introduced the notations
 \bea
\langle G_1^{(n)} \rangle\,&=&
2f_n-\frac{{s}_n^2}{1+\xi^2}\,g_n\,,
 \label{48}\\
 \langle G_3^{(n)} \rangle&=&
\left[\left(\frac{1}{1-y}+1-y\right)f_n-
\left(1+\frac{y^2}{1-y}\right)\frac{{s_n^2}}{1+\xi^2}g_n \right]
\zeta_3+ \frac{y(2-y){c}_n}{1-y}\,h_n\,P_c\,.
 \nn
 \eea

At last, the correlation of the final particles' polarizations are
 \bea
H_{11}^{(n)}&=&\frac{ys_n}{\sqrt{1+\xi^2}}h_n P_c +\frac{y}{1-y}
\left[(2-y)(f_n-g_n)+(1-y+\Delta)s_n^2g_n\right]\zeta_2
-\frac{yc_n s_n}{\sqrt{1+\xi^2}}g_n \zeta_3 \, ,
 \nonumber \\
H_{21}^{(n)}&=&-\frac{y}{1-y}
\left\{(2-y)(f_n-g_n)+\left[1+(1-y)\Delta\right]
s_n^2g_n\right\}\zeta_1 \, ,
 \nonumber \\
H_{31}^{(n)}&=& \frac{y c_n s_n}{(1-y)\sqrt{1+\xi^2}}g_n \zeta_1
\, , \ \ \ H_{12}^{(n)}= 2 c_n h_n P_c\, \zeta_1 \, ,
 \label{49} \\
H_{22}^{(n)}&=&-\frac{y c_n s_n}{(1-y)\sqrt{1+\xi^2}}\,g_n+ 2 c_n
h_n P_c\, \zeta_2 -\frac{y s_n}{(1-y)\sqrt{1+\xi^2}}\,h_n P_c\,
\zeta_3 \, ,
 \nonumber \\
H_{32}^{(n)}&=&\frac{y}{1-y}
\left[(2-y)f_n-\frac{s_n^2}{1+\xi^2}g_n\right]+\frac{y s_n
}{\sqrt{1+\xi^2}}\,h_n P_c\, \zeta_2+\frac{2-2y+y^2}{1-y}\,c_n h_n
P_c\, \zeta_3 \, ,
 \nonumber \\
H_{13}^{(n)}&=&\left[\frac{2-2y+y^2}{1-y}(f_n-g_n)
+\left(1+\frac{1-y+y^2}{1-y}\Delta\right)s_n^2\,g_n\right]\,\zeta_1
\,,
 \nonumber \\
H_{23}^{(n)}&=&-\frac{y s_n}{\sqrt{1+\xi^2}}\,h_n
P_c+\left[\frac{2-2y+y^2}{1-y}(f_n-g_n) +\left(\frac{1-y+y^2}{1-y}
+ \Delta \right)\,s_n^2\,g_n\right]\,\zeta_2+
 \nn\\
&& \frac{y c_n s_n}{\sqrt{1+\xi^2}}\,g_n \,\zeta_3 \, ,
 \nonumber \\
H_{33}^{(n)}&=&-\frac{y c_n s_n}{(1-y)\sqrt{1+\xi^2}}\,g_n\,
\zeta_2 + \left[2(f_n-g_n)+(1+\Delta)\,s_n^2\,g_n\right]\, \zeta_3
\, .
 \nn
 \eea

\section{Going to the collider system}

\subsection{Exact relations}

As an example of the application of the above formulas, let us
consider the nonlinear Compton scattering in the frame of
reference, in which an electron performs a head-on collisions with
a laser photons, i.e. in which ${\bf p} \,\parallel \,(- {\bf
k})$. In what follows, we call it as the ``collider system''. In
this system we choose the $z$-axis along the initial electron
momentum ${\bf p}$. Azimuthal angles $\varphi,\; \beta$ and
$\beta'$ of vectors ${\bf k}',\; \mbox{\boldmath$\zeta$}$ and
$\mbox{\boldmath$\zeta$}'$ are defined with respect to one fixed
$x$-axis.

In such a system the unit vectors ${\bf n}_j$, defined in
(\ref{28}), has a simple form:
\begin{equation}
{\bf n}_1={{\bf p}\times {\bf p}'\over |{\bf p}\times {\bf p}'|}=
{{\bf k}\times {\bf k}'\over |{\bf k}\times {\bf k}'|}\,,\;\;\;
{\bf n}_2={{\bf p}\times {\bf n}_1\over |{\bf p}\times{\bf n}_1|}=
{{\bf k}'_{\perp} \over |{\bf k}'_{\perp}|}\,,\;\; \; {\bf
n}_3={{\bf p}\over |{\bf p}|}\,,
 \label{50}
\end{equation}
and invariants $\zeta_j$ are equal to
 \be
\zeta_1 =\mbox{\boldmath$\zeta$}{\bf n}_1 =\zeta_{\perp}
\sin{(\varphi-\beta)}\,,\;\;\; \zeta_2 =
\mbox{\boldmath$\zeta$}{\bf n}_2 =\zeta_{\perp}
\cos{(\varphi-\beta)} \,,\;\; \zeta_3 =
\mbox{\boldmath$\zeta$}{\bf n}_3 =2\lambda_e\,.
 \label{51}
\end{equation}
Here ${\bf k}'_{\perp}$ and $\mbox{\boldmath$\zeta$} _{\perp}$
stands for the transverse components of the vectors ${\bf k}'$ and
$\mbox{\boldmath $\zeta$}$ with respect to the vector ${\bf p}$,
and $\zeta_{\perp} = |\mbox{\boldmath$\zeta$} _{\perp}|$.
Therefore, in this frame of reference, $\zeta_1$ is the transverse
polarization of the initial electron perpendicular to the
scattering plane, $\zeta_2$ is the transverse polarization in that
plane and $\zeta_3$ is the doubled mean helicity of the initial
electron.

A phase volume element in (\ref{33}) is equal to
\begin{equation}
d\Gamma=dy\;d\varphi\,,
 \label{52}
\end{equation}
and the differential cross section, summed over spin states of the
final particles, is
 \be
{d\sigma_n\over dy \, d\varphi} = {r_e^2\over x}\; F_0^{(n)} \,.
 \label{53}
 \ee
Integrating this expression over $\varphi$, we find that (similar
to the linear Compton scattering) the differential $d\sigma_n /
dy$ and the total cross sections $\sigma_n$ for a given harmonic
$n$ do not depend on the the transverse polarization of the
initial electron  :
 \bea
&&{d\sigma_n\over dy} = {2\pi r_e^2\over x}\, \langle F_0^{(n)}
\rangle \,,\;\; \sigma_n ={2\pi r_e^2\over x}\, \int_0^{y_n}
\langle F_0^{(n)} \rangle \, dy\,,
 \label{54}
 \\
&&\langle F_0^{(n)} \rangle= \left({1\over 1-y}+1-y\right)\,f_n-
{s_n^2\over 1+\xi^2}\, g_n +{y(2-y)\over 1-y}\, c_n
\,h_n\,\zeta_3\,P_c \,.
 \nn
 \eea

The polarization of the final photon is given by Eqs. (\ref{36}).
The polarization vector of the final electron is determined by Eq.
(\ref{38}), but unfortunately, the unit vectors ${\bf n}'_j$ in
this equation has no simple form similar to (\ref{50}). Therefore,
we have to find the characteristics that are usually used for
description of the electron polarization --- the helicity of the
final electron $\lambda_e'$ and its transverse (to the vector
${\bf p}'$) polarization $\mbox{ \boldmath $\zeta$}^{\prime}_\bot$
in the considered collider system. For this purpose we introduce
unit vectors $\mbox{ \boldmath$\nu$}_j$, one of them is directed
along the momentum of the final electron ${\bf p}^{\prime}$ and
two others are in the transverse plane (compare with Eq.
(\ref{50})):
\begin{equation}
\mbox{\boldmath$\nu$}_1 = {\bf n}_1, \;\;\;
\mbox{\boldmath$\nu$}_2 ={{\bf p}'\times {\bf n}_1 \over|{\bf
p}'\times{\bf n}_1|}, \;\;\; \mbox{\boldmath$\nu$}_3 = {{\bf
p}'\over |{\bf p}'|};\;\;\; \mbox{\boldmath$\nu$}_i \,
\mbox{\boldmath$\nu$}_j = \delta_{ij}\,.
 \label{55}
\end{equation}
After that one has
 \be
\mbox{\boldmath$\zeta$}'\,\mbox{\boldmath$\nu$}_1=
\zeta'_{\perp}\sin{(\varphi-\beta')}\,,\;\;
\mbox{\boldmath$\zeta$}'\,\mbox{\boldmath$\nu$}_2=
\zeta'_{\perp}\cos{(\varphi-\beta')}\,,\;\;
\mbox{\boldmath$\zeta$}'\,\mbox{\boldmath$\nu$}_3= 2\lambda'_e\,.
 \label{56}
 \ee
Therefore, $\mbox{\boldmath$\zeta$}'\,\mbox{\boldmath$\nu$}_1$ is
the transverse polarization of the final electron perpendicular to
the scattering plane,
$\mbox{\boldmath$\zeta$}'\,\mbox{\boldmath$\nu$}_2$ is the
transverse polarization in that plane and
$\mbox{\boldmath$\zeta$}'\,\mbox{\boldmath$\nu$}_3$ is the doubled
mean helicity of the final electron.

Let us discuss the relation between the projection
$\mbox{\boldmath$\zeta$}'\,\mbox{\boldmath$\nu$}_j$, defined
above, and the invariants $\zeta_j'$, defined in (\ref{26}). In
the collider system the vectors $\mbox{\boldmath$\nu$}_1$ and
${\bf n}_1'$ coincide
 \be
\mbox{\boldmath$\nu$}_1={\bf n}_1'\,,
 \ee
therefore,
 \be
\mbox{\boldmath$\zeta$}'\,\mbox{\boldmath$\nu$}_1 = \zeta_1'\,.
 \ee
Two other unit vectors ${\bf n}_2'$ and ${\bf n}_3'$ are in the
scattering plane and they can be obtained from vectors
$\mbox{\boldmath$\nu$}_2$ and $\mbox{\boldmath$\nu$}_3$ by the
rotation around the axis $\mbox{\boldmath$\nu$}_1$ on the angle
$(-\Delta \theta)$:
 \be
\mbox{\boldmath$\zeta$}'\,\mbox{\boldmath$\nu$}_2=
\zeta'_2\,\cos{\Delta \theta}+ \zeta'_3\,\sin{\Delta \theta}\,
,\;\; \mbox{\boldmath$\zeta$}'\,\mbox{\boldmath$\nu$}_3=
\zeta'_3\,\cos{\Delta \theta}- \zeta'_2\,\sin{\Delta \theta}\,,
 \label{59}
 \ee
where
 \be
\cos{\Delta \theta}= \mbox{\boldmath$\nu$}_3{\bf n}_3'\,\;\;
\sin{\Delta \theta}= - \mbox{\boldmath$\nu$}_3{\bf n}_2'\,.
 \ee
As a result, the polarization vector of the final electron
(\ref{38}) is expressed as follows
 \be
\mbox{\boldmath$\zeta$}^{(f)}= \mbox{\boldmath$\nu$}_1\,{G_1\over
F_0}\,+ \mbox{\boldmath$\nu$}_2 \left( {G_2\over F_0} \cos{\Delta
\theta} + {G_3\over F_0} \sin{\Delta \theta} \right) \,+
\mbox{\boldmath$\nu$}_3 \left( {G_3\over F_0} \cos{\Delta \theta}
- {G_2\over F_0} \sin{\Delta \theta} \right)\,.
 \label{61}
 \ee

\subsection{Approximate formulas}

All the above formulas are exact. In  this subsection we give some
approximate formulas useful for application to the important case
of high-energy $\gamma \gamma$ and $\gamma e$ colliders. It is
expected that in the conversion region of these colliders, an
electron with the energy $E\sim 100$ GeV performs a head--on
collisions with a laser photons having the energy $\omega \sim 1$
eV per a single photon~\cite{TESLA}. In this case the most
important kinematical range corresponds to almost back-scattered
final photons, i. e. the initial electron is ultra-relativistic
and the final photon is emitted at small angle $\theta_\gamma$
with respect to the $z$-axis:
\begin{equation}
E \gg m, \;\;\; \theta_\gamma \ll 1.
 \label{62}
\end{equation}
In this approximation we have
\begin{equation}
x\approx {4E\omega\over m^2}\,,\;\;\;\; y\approx{\omega'\over E}
\approx 1- {E'\over E}\,,
 \label{63}
\end{equation}
therefore, Eq. (\ref{53}) gives us the distribution of the final
photons over the energy and the azimuthal angle. Besides, the
photon emission angle for reaction (\ref{2}) is
 \be
 \theta_\gamma \approx {m\over E}\, \sqrt{nx+1+\xi^2}\;\sqrt{{y_n\over
 y}-1}
 \label{64}
 \ee
and $\theta_\gamma \to 0$ at $y\to y_n$. For a given $y$, the
photon emission angle increases with the increase of $n$. The
electron scattering angle is small
\begin {equation}
\theta_e \approx {y\theta_\gamma\over 1-y}\leq {2n\, \over
\sqrt{1+\xi^2}}\, {\omega\over m}\, .
 \label{65}
\end {equation}

It is nor difficult to check that, for the considered case, the
angle $\Delta \theta$ between vector $\mbox{\boldmath$\nu$}_3$ and
${\bf n}'_3$ is very small:
 \be
\Delta \theta \approx |\mbox{\boldmath$\nu$}_{3\perp}-{\bf
n}'_{3\perp}| \approx {m \theta_e \over 2 E'} \leq {n\over
\sqrt{1+\xi^2}}\,{\omega\over E'}\, .
 \ee
It means that invariants $\zeta_j'$ from (\ref{26}) almost
coincide with projections
$\mbox{\boldmath$\zeta$}'\,\mbox{\boldmath$\nu$}_j$, defined in
(\ref{56}), and that the exact equation (\ref{61}) for the
polarization of the final electron can be replaced  with a high
accuracy by the approximate equation
 \be
\mbox{\boldmath$\zeta$}^{(f)} \approx \sum_{j=1}^3\,{G_j\over
F_0}\, \mbox{\boldmath$\nu$}_j \,.
 \label{67}
 \ee

\section{Summary and comparison with other papers}

Our main result is given by Eqs. (\ref{45})--(\ref{49}) which
present 16 functions $F_0,\; F_j,\; G_j$ and $H_{ij}$ with $i,\,j
= 1\div 3$. They describe completely all polarization properties
of the nonlinear Compton scattering in a rather compact form.

In the literature we found 5 functions which can be compared with
ours $F_0, F_2$ and $G_j$. Functions $F_0$ and $F_j$ enter the
total cross section (\ref{35}), (\ref{54}), the differential cross
sections (\ref{53}), (\ref{54}) and the Stokes parameters of the
final photons (\ref{36}). Two of these functions were calculated
in~\cite{NR-review} (the function $F_0$ only at $\zeta_j=0$),
in~\cite{GS,Tsai,Yokoya} (the functions $F_0$ and $F_2$ only at
$\zeta_1=\zeta_2=0$) and in~\cite{BPM} (the function $F_0$ for
arbitrary $\zeta_j$). For these cases our results coincide with
the above mentioned ones.

The polarization of the final electrons is described by functions
$G_j$ (\ref{47}), (\ref{48}). They enter the polarization vector
$\mbox{\boldmath$\zeta$}^{(f)}$ given by exact (\ref{38}),
(\ref{61}) and approximate (\ref{67}) equations. Here our results
differ slightly from those in~\cite{Yokoya,BPM}. Namely, function
$G_2$ at $\zeta_1=\zeta_2=0$ given in~\cite{Yokoya} and functions
$G_j$ for arbitrary $\zeta_j$ obtained in~\cite{BPM} coincide with
ours. However, the polarization vector
$\mbox{\boldmath$\zeta$}^{(f)}$ in the collider system is obtained
in papers~\cite{Yokoya,BPM} only in the approximate form
equivalent to our approximate Eq. (\ref{67}).

The correlation of the final particles' polarizations are
described by functions $H_{ij}$ given in Eqs. (\ref{49}). We did
not find in the literature any on these functions.

At small $\xi^2$ all harmonics with $n > 1$ disappear due to
properties (\ref{42}) and (\ref{44}),
 \be
d\sigma_n(\xi^{\prime}_i,\zeta^{\prime}_j) \propto \xi^{2(n-1)}
\;\; \mbox{ at } \;\; \xi^2\to 0\,.
 \label{c1}
 \ee
We checked that in this limit our expression for $d\sigma
(\xi^{\prime}_i,\zeta^{\prime}_j)$ coincides with the result known
for the linear Compton effect, see Appendix.

\section*{Acknowledgements}

We are grateful to I.~Ginzburg, M.~Galynskii, A.~Milshtein,
S.~Polityko and V.~Telnov for useful discussions. This work is
partly supported by INTAS (code 00-00679) and RFBR (code
03-02-17734); D.Yu.I. acknowledges the support of Alexander von
Humboldt Foundation.

\section*{Appendix: Limit of the weak laser field}

At $\xi^2\to 0$, the cross section (\ref{33}) has the form
\begin{equation}
d\sigma (\xi^{\prime}_i,\zeta^{\prime}_j) = {r_e^2\over 4x}\;F
\;d\Gamma;\;\;\;d\Gamma = \delta (p+k-p-k')\;{d^3k'\over
\omega'}{d^3p'\over E'}\,,
 \label{a1}
 \end{equation}
where
\begin{equation}
F=F_0+\sum ^3_{j=1}\left( F_j\xi'_j\; + \;G_j
\zeta^{\prime}_j\right) +
\sum^3_{i,j=1}H_{ij}\,\zeta^{\prime}_i\,\xi^{\prime}_j \,.
 \label{a2}
\end{equation}
To compare it with the cross section for the linear Compton
scattering, we should take into account that the Stokes parameters
of the initial photon have values (\ref{20}) and that our
invariants $c_1$ and $s_1$ transforms at $\xi^2=0$ to
 \be
c=1- 2r\,,\;\; s =2\sqrt{r(1-r)}
 \label{a3}
 \ee
with $r={y/[ x(1-y)]}$. Our functions
 \bea
F_0&=&{1\over 1-y}+1-y-s^2-y\left(s\, \zeta_2 -{2-y\over 1-y}c\,
\zeta_3\right) P_c\,,\;\; F_1={y\over 1-y} s \,P_c\, \zeta_1\,,
 \label{a4}
 \\
F_2&=&\left({1\over 1-y} +1-y\right)\, c\,P_c- {ys c}\,\zeta_2+
 y\left( {2-y\over 1-y}\,- s^2\right) \zeta_3\,,
 \nn
 \\
F_3&=& s^2  -{y\over 1-y} s\, P_c\, \zeta_2
 \nn
 \eea
coincide with those in~\cite{GKPS83}. Our functions
 \bea
G_1&=&(1+c^2)\,\zeta_1\,,\;\; G_2=-{ys\over 1-y}\, P_c + (1+
c^2)\, \zeta_2 - {ysc \over 1-y} \,\zeta_3\,,
 \label{a5}
  \\
G_3 &=& y\, {2-y \over 1-y}\,c\, P_c +  ysc \, \zeta_2+
 \left(1+  {1-y+y^2\over 1-y} \,c^2  \right) \, \zeta_3\,.
 \nn
 \eea
coincide with functions $\Phi_j$ given by Eqs. (31)
in~\cite{KPS98}. At last, we check that our functions
 \bea
H_{11}&=&ys P_c +ys^2\zeta_2 -y c s \zeta_3 \, , \;\;
H_{21}=-\frac{y}{1-y} s^2\zeta_1 \, ,
 \nonumber \\
H_{31}&=& \frac{y c s}{1-y}\, \zeta_1 \, , \;\; H_{12}= 2 c P_c\,
\zeta_1 \, ,
 \label{a6} \\
H_{22}&=&-\frac{y c s}{1-y} + 2 c \,P_c \,\zeta_2 -\frac{y s
}{1-y} P_c\, \zeta_3 \, ,
 \nonumber \\
H_{32}&=&\frac{y}{1-y} \left(2-y-s^2\right)+y s\, P_c\,
 \zeta_2+\frac{2-2y+y^2}{1-y}c \, P_c\, \zeta_3 \, ,
 \nonumber\\
H_{13}&= & s^2\,\zeta_1 \,,\;\;  H_{23}=- ys P_c+
\frac{1-y+y^2}{1-y} s^2\, \zeta_2+ y c s \,\zeta_3 \,,
 \nonumber \\
H_{33}&=&-\frac{y c s}{1-y}\,  \zeta_2 +
 s^2\,\zeta_3
 \nn
 \eea
coincide with the corresponding functions from \cite{Grozin}. At
such a comparison, one needs to take into account that the set of
unit 4-vectors, used in our paper (see Eqs. (\ref{19}),
(\ref{23})), and one used in paper \cite{Grozin} are different.
The relations between our notations and the notations used
in~\cite{Grozin}, which are marked below by super-index $G$, are
the following:
 \bea
\xi_{1,3}^G &=& -\xi_{1,3}\,,\;\; \xi_{2}^G = \xi_{2}\,,\;\;
\xi_{1,3}^{'G} = -\xi'_{1,3}\,,\;\; \xi_{2}^{'G} = \xi'_{2}\,,
 \label{aa}
 \\
\zeta_{1,3}^G& =&\zeta_{3,1}\,,\;\; \zeta_2^G=-\zeta_2\,,\;\;
\zeta_{1}^{'G} =c\zeta_{3}'- s \zeta_2'\,,\;\; \zeta_{2}^{'G} =
-c\zeta_{2}'- s \zeta_3'\,,\;\; \zeta_3^{'G} =\zeta_1'\,.
 \nn
  \eea

\end{document}